\documentclass[letterpaper,10pt]{article} 

\usepackage{opticameet3} 

\usepackage{titlesec}
\titlespacing\section{0pt}{4pt plus 2pt minus 2pt}{4pt plus 2pt minus 2pt}

\newcommand\authormark[1]{\textsuperscript{#1}}

\usepackage{amsmath,amssymb}
\usepackage[colorlinks=true,bookmarks=false,citecolor=blue,urlcolor=blue]{hyperref} 

\begin{document}

\title{Classical Attack on Bell Inequalities}


\author{Aishi Guha,\authormark{1} Noah A. Davis,\authormark{2} and Brian R. La Cour\authormark{2,*}}

\address{\authormark{1} The University of Texas at Austin, Austin, TX 78712\\
\authormark{2}Applied Research Laboratories, The University of Texas at Austin, P.O. Box 8029, Austin, TX 78713-8029}

\email{\authormark{*}blacour@arlut.utexas.edu} 

\begin{abstract}
Representing multi-mode squeezed light with a Gaussian random vector, our locally deterministic detection model challenges the CHSH game, achieving fidelities exceeding 96\%. Squeezing strength, detector threshold, and efficiency influence the security of the quantum bound.

\end{abstract}

\section{Introduction}
Entanglement swapping has been demonstrated at high fidelity (over 80\%) \cite{Humphreys2018}, but quantum fidelity alone is not adequate to ensure security; one must also consider efficiency \cite{Larsson2014}, as prior work has shown that local hidden-variable models can violate the Clauser-Horne-Shimony-Holt (CHSH) inequality for low detection efficiencies \cite{Gisin1999}.

We provide a physics-based approach to modeling quantum optical phenomena and apply it to an entanglement swapping experiment, examining the fidelity of the resulting state and the efficiency of the generation and analysis process. Our model follows prior work in modeling quantum optics using a reified vacuum field and a classical, threshold-based detector model \cite{LaCour2020,LaCour2021a,LaCour2023}.  In this model, entanglement swapping relies on the post-selection of successful partial Bell-state measurements, a point that has been emphasized previously \cite{Broadbent2007}. Such models help delineate the quantum-classical boundary and provide physical insight for designing future quantum networks.

We perform full QST on our model and directly assess the fidelity of the generated state as a function of the squeezing strength and detector thresholds.  We also examine the efficiency of QST and successful partial Bell state measurements.  We emphasize that for successful entanglement swapping the efficiency of the entire process, from initial state generation to final measurement and analysis, must be considered to guard against spoofing.


\section{Model Description}
We consider an optical experiment using parametric downconversion (PDC) sources, beam splitters, polarizers, and detectors. A PDC source prepares a Gaussian multimode squeezed state described by a pair of random Jones vectors.  This admits a Gaussian Wigner function and a corresponding probability density function.  

The form of the Jones vectors derives from the Bogoliubov equations for a multimode squeezed state, replacing the annihilation operator for each mode with the corresponding random variables \cite{LaCour2021a}. This defines exactly a macroscopic Bell state corresponding to $\vert\Psi^{-}\rangle = \frac{1}{\sqrt{2}} \Bigl[ \vert HV\rangle - \vert VH\rangle \Bigr] $, as described in \cite{Iskhakov2011}. 

We adopt a model of photon detection in which an event occurs when the amplitude of the incident wave exceeds some specified threshold \cite{LaCour2020,Khrennikov2014}.  In particular, we say a successful entanglement swapping event occurs when the Bell state measurement threshold $\gamma_B$ is exceeded in exactly one detector in each spatial mode. Though Alice and Bob's PDC sources are independent, post-selection on successful measurements creates a statistical dependence between them. This provides a simple and intuitive classical picture for understanding entanglement swapping. Note that we use two additional detectors each for Alice and Bob with detection threshold $\gamma_A$. A valid detection event for Alice is one in which a detection occurs on exactly one of the two detectors.


\section{Experiment and Results}
The numerical experiment is performed as follows: First, complex Gaussian random variables are used to realize Alice and Bob's variables $\vec{a}_{1}$, $\vec{a}_{2}$, $\vec{b}_{1}$ and $\vec{b}_{2}$. Next, $\vec{a}_{1}$ and $\vec{b}_{1}$ are combined through the beam splitter transformations and then filtered through polarizers. We say that a successful BSM event has occurred if the resulting variables both satisfy the Bell-state measurement detector threshold. For successful BSM events, the variables $\vec{a}_{2}$ and $\vec{b}_{2}$ are subject to quantum state tomography, and the density matrix is reconstructed. Ten trials were conducted for each combination of the parameters $r$, $\gamma_A$, and $\gamma_B$ (squeezing strength, BSM detection threshold, and QST detector threshold), and each trial included at least 10,000 successful BSM events.



Bell-state measurement efficiency of unity is achieved only for the trivial BSM detector threshold $\gamma_B = 0$, and efficiency falls off as this threshold is increased. Similarly, the BSM efficiency falls to zero as the squeezing strength decreases.  The transition from high to low efficiency occurs roughly where $r = 1$ and $\gamma_B = 1$.

\begin{figure}
    \vspace{-5ex}
   \begin{minipage}{0.48\textwidth}
     \centering
     \includegraphics[width=.9\linewidth, height = 4cm]{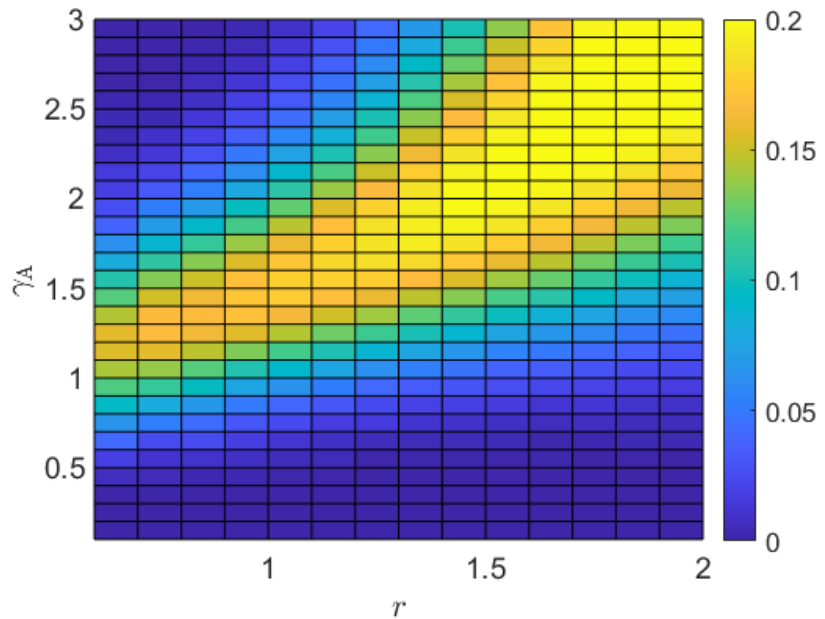} 
     \caption{QST efficiency for BSM threshold $\gamma_B = 1.2$.}
     \label{fig:efficiencyvaryingr.jpg}
   \end{minipage}
   \begin{minipage}{0.48\textwidth}
     \centering
     \includegraphics[width=.8\linewidth, height = 4cm]{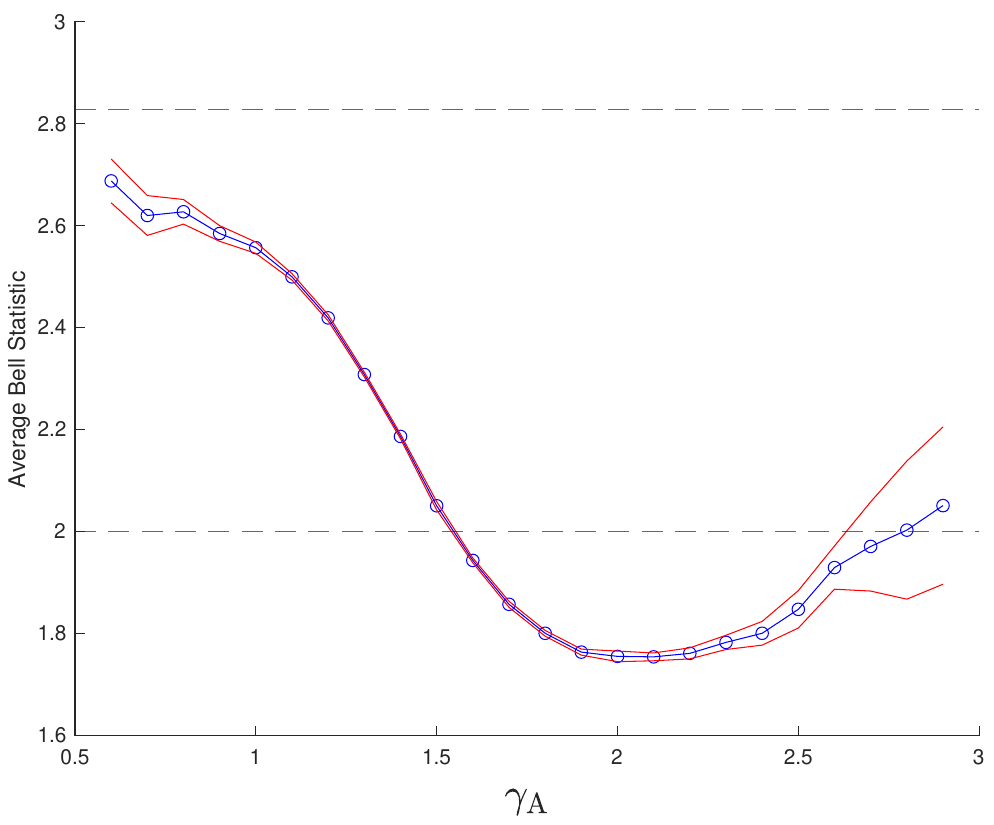} 
     \caption{CHSH score for $r = 0.9$ and $\gamma_B = 2.3$.}
     \label{fig:g2-2.3_r-1_Bd.jpg}
   \end{minipage}
    \vspace{-4ex}
\end{figure}

The QST efficiency, by contrast, measures the proportion of successful BSM events that also result in a valid coincident detection for Alice and Bob. The dependence of QST efficiency on squeezing strength is displayed in Fig. \ref{fig:efficiencyvaryingr.jpg} for a fixed BSM threshold of $\gamma_B = 1.2$.  There is a strong dependence on $r$, with higher values resulting in a brighter entanglement source and a higher efficiency when the QST threshold $\gamma_A$ is suitably matched.

Maximum fidelity of $F = 0.964 \pm 0.009$ is achieved at the parameter values of $r = 0.9$, $\gamma_B = 2.3$, and $\gamma_A = 0.6$.


\section{Discussion}
The Bell state measurement efficiency specifically refers to the fraction of random realizations that result in successful BSM events. Interestingly, since only successful BSM events are considered in the reconstruction of the state, BSM efficiency can be considered a measure of the influence of the detection loophole on the experiment.

The quantum state tomography efficiency refers to the rate of successful QST measurements, given that a successful BSM event has occurred. In successful QST measurement Alice and Bob each record exactly one detection; low QST thresholds $\gamma_A$ result in multiple detections, and high thresholds prevent even single detections.

We see a general trend of higher fidelities corresponding to higher BSM threshold $\gamma_B$ and lower QST threshold $\gamma_A$.  Fidelity has a more complicated relationship to squeezing strength, however, taking maximum values with a squeezing strength of $r \approx 1$. Perhaps most experimentally relevant, we reach a fidelity of $F=0.81 \pm 0.01$ for $\gamma_A=\gamma_B=2.3$ and $r=0.6$, approximately matching or exceeding experimental fidelities \cite{JinTakeokaEtAl2015,Sun2017}


The CHSH score gives another measure on which to judge the quality of our reconstructed states. Fig.\;\ref{fig:g2-2.3_r-1_Bd.jpg} shows that the score surpasses the classical limit of $S=2$ for a range of thresholds, with a maximum of $S=2.69 \pm 0.04$. We post-select on successful BSM events,  taking advantage of the detection loophole to violate the classical bound and also beating previous experimental results \cite{KaltenBaekZeilingerEtAl2009} without exceeding the Tsirelson bound of $2\sqrt{2}$.


\section{Conclusion}

We present a model of optical entanglement swapping using a classical, stochastic approach, allowing fidelities of up to $0.963\pm 0.009$ and violations of the classical bound on the CHSH inequality with scores up to $2.69\pm 0.04$. These results can be understood as a consequence of the detection loophole and the low efficiency of the quantum state tomography measurements. The model allows a reliable spoof of quantum entanglement experiments when the method of verification does not adequately account for detection efficiency across the entire measurement process. This work provides a path for future stochastic modeling of entangled networks and may be useful in predicting relative, qualitative parameter dependencies for actual entanglement swapping experiments.


\section{Acknowledgments}

This work was funded by the Office of Naval Research (N00014-18-1-2107,N00014-23-1-2115), the National Science Foundation (1842086), and the Freshman Research Initiative (FRI) at The University of Texas at Austin.



\end{document}